\documentclass[floatfix,reprint,amsmath,amssymb,aps,prl,showpacs,showkeys]{revtex4-1}
\usepackage{fullpage,amssymb,times}

\usepackage{graphicx}% Include figure files
\usepackage{dcolumn}% Align table columns on decimal point
\usepackage{bm}% bold math

\newcommand{\blot}[1]{}

\begin{document}

\preprint{GKW-2811}

\title{Bak--Sneppen type models and rank-driven processes}

\author{Michael Grinfeld}
\email{m.grinfeld@strath.ac.uk}
\author{Philip A. Knight}
\email{p.a.knight@strath.ac.uk}
\author{Andrew R. Wade}
\email{andrew.wade@strath.ac.uk}
\affiliation{Department of Mathematics and Statistics\\
University of Strathclyde,\\
26 Richmond Street, Glasgow G1 1XH, UK}

\date{\today}

\begin{abstract}
  \noindent The Bak--Sneppen model is a simple stochastic
  model of evolution that exhibits self-organized criticality and for which few
  analytical results have been established. In the
  original Bak--Sneppen model and many subsequent variants,
  interactions among the evolving species are tied to a specified topology.
   We report a surprising connection between Bak--Sneppen
  type models and more tractable Markov processes that evolve
  without any reference to an underlying topology. Specifically, we show
  that in the case of a large number of species, the long time
  behaviour of the fitness profile in the anisotropic Bak--Sneppen model can be
  replicated by a model with a purely rank-based update rule whose
  asymptotics can be studied rigorously.
\end{abstract}


\maketitle

In \cite{BS}, Bak and Sneppen introduced a very fruitful and simple
model of evolution that exhibits interesting dynamics but has proved
surprisingly hard to analyse. The classical Bak--Sneppen (BS) model is
a simple stochastic coarse-grained model of evolution of an ecosystem
consisting of a fixed number $N$ of evolutionary niches organised in a ring.  Each niche is occupied by a
species with a particular {\em fitness} value in $[0,1]$.
Direct
inter-species interactions (predation, competition, etc.) occur only
between species in neighbouring niches.
The dynamics
of the system is driven by the removal (extinction) of the least fit
species in the entire system, whose niche is taken over by a new
species; the extinction of the least fit species induces changes in the
fitnesses of the species in the two neighbouring niches.  In
this letter, we show that a process whose update rule is defined
solely in terms of the {\em ranks} of the fitness values, without any
reference to topology of interactions, exhibits asymptotic behaviour
similar to the (anisotropic version of the) BS model as well as
self-organized criticality \cite{jensen}. We call processes of this
type {\em rank-driven processes} (RDPs) and analyse them rigorously
in~\cite{GKW11}. RDPs are of independent mathematical interest and can
be used to define new evolution models. Despite the considerable
impact of the Bak--Sneppen model on the physics community and beyond
[\onlinecite{BS1}], so far only a small number of rigorous results on the BS model
have  been obtained, such as those of Meester and Znamenski
\cite{MZ} on the non-triviality of the steady-state distribution. In
this letter, by exploiting the tools for analysis of RDP models
developed in~\cite{GKW11}, we provide an approach for establishing new
results in this active area.

The BS model \cite{BS} is a discrete-time Markov process which
advances every time there is a species extinction event. Each
species occupying the $N$ niches is initially assigned a fitness $x_k
\in [0,1]$ ($k \in \{1,\ldots,N\}$) chosen independently from the
uniform distribution on the unit interval, $U[0,1]$. At each step of
the algorithm, we choose the smallest of all the $x_k$, $x_{kmin}$
say, and replace $x_{kmin}$ and its two nearest neighbours $x_{kmin
  \pm 1}$ (indices calculated modulo $N$) by new independent $U[0,1]$
random numbers. In simulations with large $N$, the marginal
distribution of the fitness at any particular niche is seen to evolve
to a $U[s^*,1]$ distribution, with $s^* \approx 0.667$.

A number of variants of Bak and Sneppen's original model have been
introduced which evolve according to different topological
criteria. One simple variant is the {\em anisotropic} Bak--Sneppen
(aBS) model, in which, in addition to the least fit species, only its {\em
  right-hand} nearest neighbour is replaced. The aBS model is the main
focus of this letter because while it simplifies calculations, it
preserves the key qualitative phenomena of the original BS model. For
example, the aBS model also gives rise (according to large-$N$
simulations) to a threshold value $s^* \approx 0.724$ \cite{GD}.
One contribution of the present letter is to propose a characterization
for $s^*$ in terms of ostensibly simpler quantities associated with aBS.
Our
arguments connecting rank-driven processes to aBS apply to BS too.

Another variant on the BS model which eliminates topology is the
mean-field version analysed in \cite{FSB, deBoer, LP}, in which one
replaces the smallest fitness and $K-1$ randomly chosen other ones.
Below we show that such models fall within the RDP framework.

Consider a process in which at each update the species with the
smallest fitness and the $R$-th ranked fitness are replaced, where $R$
is a random variable on $\{2,3,\ldots,N\}$ ($R=N$ corresponding to the
largest fitness) sampled independently from a distribution $P [ R = k]
= f_N (k)$ where $f_N (k) \geq 0$ and $\sum_{k=2}^N f_N(k) = 1$. This
is an example of a rank-driven process.  The complexity of this
RDP is intermediate between that of the aBS model and the mean-field
model of \cite{FSB, deBoer, LP}; the latter is the special case of a
RDP with $f_N(k) = \frac{1}{N-1}$ for all $k \in \{2,\ldots,N\}$.  The
RDP has the advantage over aBS that it can be analysed rigorously. We
have strong numerical evidence that for a judicious choice of $f_N(k)$
this simpler model can replicate the asymptotic behaviour of aBS.

Specifically, one can choose $f_N(k)$ to be $f^{\rm aBS}_N(k)$, the
{\em empirical} distribution of the rank of the second chosen site in
aBS: if we let $P(k,M)$ be the number of times the $k$-th ranked
element, $k \geq 2$, is the right neighbour of the smallest element in
$M$ iterations of the aBS algorithm,
\begin{equation}
\label{aBSf}
f^{\rm aBS}_N(k) = \lim_{M\rightarrow \infty}
\frac{1}{M} P(k,M).
\end{equation}
Heuristically, we expect that given suitable ergodicity properties for
the aBS Markov process on the uncountable state space $[0,1]^N$, this
limit will exist with probability one.

\begin{figure}[h!t]
\begin{center} \includegraphics[height=4.5cm]{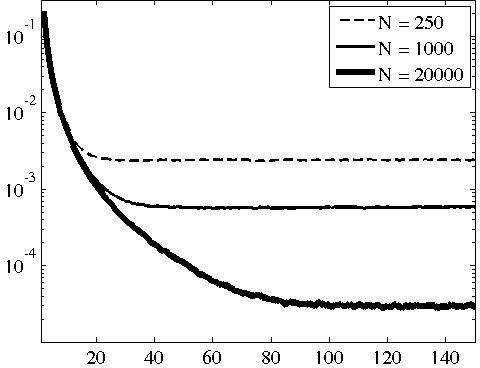} \end{center}
\vspace*{-.8cm}
\caption{Plot of $f^{\rm aBS}_N(k)$, $k \in \{2,\,\ldots,\, 150\}$ for $N=250,
1000,
20000$.}
\label{fig1}
\end{figure}

Little is known analytically about $f^{\rm aBS}_N$, due to the
difficulty of the aBS model, but
$f^{\rm aBS}_N$ can be accurately numerically computed.  Figure
\ref{fig1} shows simulation estimates of $f^{\rm aBS}_N(k)$ for small values of $k$
and different values of $N$.

The RDP with $f_N = f^{\rm aBS}_N$ computed numerically replicates
many aspects of   the dynamics of aBS. Specifically,
rigorous results show that such an RDP exhibits a (large-$N$, long-time) threshold
at some $s^*$ which is an explicit function of $f_N$:
for the particular choice $f_N = f^{\rm aBS}_N$, using our numerical
estimates for $f^{\rm aBS}_N$ gives rise to a value for $s^*$
which is very close to the value computed directly from simulations of aBS.

In \cite{GKW11}, by considering the random walk associated with
 an RDP defined in terms of $f_N$, we show that a crucial quantity is
\begin{equation}\label{eq:alpha}
\alpha = \lim_{n
\rightarrow \infty} \lim_{N \rightarrow \infty} \sum_{k=2}^n f_N(k),
\end{equation}
assuming that the $N$-limit exists. Here $\alpha \in [0,1]$ measures
the ``atomicity'' of $f_N$ as $N \rightarrow \infty$. For example, for
the mean field aBS of \cite{FSB,deBoer,LP} one has $\alpha=0$, while
if we always replace the smallest and the second smallest elements,
$\alpha=1$. The main result of \cite{GKW11} is that the threshold in
the limiting ($N \to \infty$) stationary distribution of $x$ values in
the RDP is given by
\begin{equation}\label{eq:thres}
s^*=\frac{1+\alpha}{2}.
\end{equation}

A second result of \cite{GKW11} shows that the limiting marginal
distribution at stationarity is $U[s^*,1]$, where $s^*$ is given by
(\ref{eq:thres}), provided that the selection distribution $f_N$ is
``eventually uniform'' in the sense that
\begin{equation}
\label{eq:evenu} f_N (k) \approx \frac{1-\alpha}{N}\end{equation} for
$k$ sufficiently large. This condition is satisfied with $\alpha~=~0$
for the mean-field aBS model, showing that the limiting
distribution is indeed $U[1/2,1]$ in that case, as indicated by \cite{deBoer}.

From the results of Figure \ref{fig1} we see that for a given $N$,
$f^{\rm aBS}_N(k)$ decays rapidly for small $k$ before settling down
to a uniform value.  In fact, it appears that there is a constant $C$
such that $f^{\rm aBS}_N(k) = C/N$ for large enough $k$. Thus the
numerical evidence supports the eventual uniformity condition
(\ref{eq:evenu}). Hence $\alpha = 1 - C$. Numerical results give
$\alpha \approx 0.445$ and hence $s^* \approx 0.723$, in close
agreement with the simulations of \cite{GD}. Note that $f^{\rm
  aBS}_N(2) \approx 0.209$ for all the values of $N$ in Figure
\ref{fig1}.

\begin{figure}[h!t]
\begin{center} \includegraphics[height=4.5cm]{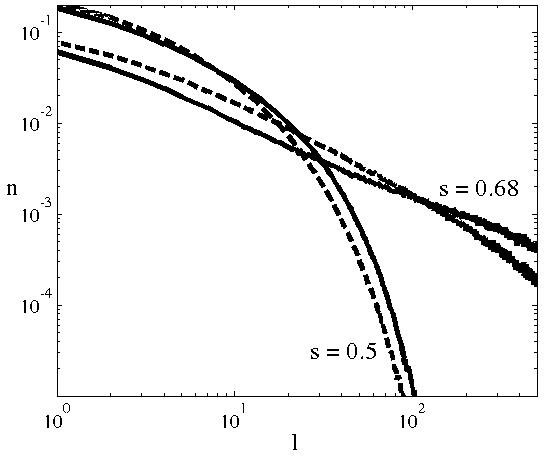} \end{center}
\vspace*{-.5cm}
\caption{Size distribution $n(\ell)$ of $s$ avalanches in aBS (solid line)
and RDP (dashed) for $s = 0.5,\, 0.68$ and $N = 1000$.}
\label{fig2}
\end{figure}

Following~\cite{BS} we define the length of an $s$-avalanche to be $\ell$
if the number of consecutive steps for which the smallest fitness
value stays below $s$ is $\ell$. We compute the distribution $n(\ell)$ of
$s$-avalanche lengths for aBS and our RDP. Representative
distributions are given in Figure \ref{fig2}. As $s$ approaches $s^*$
we find that $n(\ell)$ shows the power law behaviour characteristic of
self-organized criticality, although there is a small but clear
difference in the exponents of the two processes.

The class of RDPs defined by $f_N$ described above is contained in a wider
class of processes \cite{GKW11}, which
we now define. In this generality, an RDP is a discrete-time Markov process on the $N$-simplex
\[
\Delta_N = \{ (x_{(1)}, \ldots, x_{(N)} ) \colon
0 \leq x_{(1)} \leq \cdots \leq x_{(N)} \leq 1 \};
\]
$x_{(1)},\ldots, x_{(N)}$ are the (increasing) {\em order statistics}
of $x_1, \ldots, x_N$. The RDP evolves according to the following
Markovian rule. At each step, $K$ of the $x_k$-values are selected,
according to rank, by sampling (without replacement, according to some
specified probability distribution) from $\{1,2,\ldots,N\}$; that is,
the sample from $\{1,2,\ldots,N\}$ specifies the $x_{(k)}$ that are
chosen. The chosen $x_k$-values are replaced by new (independent)
$U[0,1]$ values. These processes are discussed in more detail in \cite{GKW11}.

The numerical evidence reported above leads to several interesting problems
for further investigation, not least the strong suggestion that the RDP with
$f_N = f_N^{\rm aBS}$ given by (\ref{aBSf}) is closely related to aBS
itself.  The exact relationship of the two processes remains to be
characterized rigorously. If one wished to define a Markov process on
$\Delta_N$ whose stationary distribution coincided with the projection onto
$\Delta_N$ of the stationary distribution of aBS, a natural candidate would
be a RDP with state-dependent selection distribution: instead of a single
$f_N(\cdot)$ one would have a family $f_N ( \, \cdot \, ; x)$ of selection
distributions conditioned on the state $x \in \Delta_N$. Thus, assuming it
exists, one would take $f_N (\, \cdot \, ; x)$ to be $f^{\rm aBS}_N (\,
\cdot \, ; x)$, the stationary distribution for aBS of the right-neighbour
of the smallest element {\em conditional} on the projection of the current
state onto $\Delta_N$ being $x$.  The fact that the numerical evidence
described above suggests that one can proceed not with a state-dependent RDP
based on $f^{\rm aBS}_N ( \, \cdot \, ; x)$ but with the simpler RDP based
on $f^{\rm aBS}_N( \, \cdot \,)$ (which is an average of the $f^{\rm aBS}_N
( \, \cdot \, ; x)$) seems to point to some important underlying property of
aBS itself. Two possible explanations are:
\begin{itemize}
\item[(a)] $f^{\rm aBS}_N( \, \cdot \,) = f^{\rm aBS}_N ( \, \cdot \, ; x)$ for
all $x$,  i.e., at stationarity there is some independence between the order
statistics and the permutation that maps sites to ranks; or
\item[(b)] $f^{\rm aBS}_N ( \, \cdot \, ; x)$ satisfies (uniformly in $x$) the
same asymptotic conditions as $f^{\rm aBS}_N( \, \cdot \,)$ that are central to
the limit behaviour, namely (\ref{eq:alpha}) and (\ref{eq:evenu}).
\end{itemize}
The stronger fact (a) would suggest that the stationary distribution of the
  RDP coincides with the projection of the stationary distribution of aBS
onto $\Delta_N$, so that the two processes share the same detailed equilibrium
properties. The weaker fact (b) would suffice to explain why the two processes
share the same threshold and characteristic $U[s^*,1]$ limit distribution.
We remark that the distributions $f^{\rm aBS}_N( \, \cdot \, ; x)$ seem
to be very difficult to evaluate numerically.

Finally, we give a very brief indication of the origin of the
threshold formula (\ref{eq:thres}); see \cite{GKW11} for
details. Consider the $s$-counting process $N_t(s)$ defined to be the
number of $x_k$-values in the interval $[0,s]$ after $t$ iterations of
the RDP defined by $f_N$. Then $N_t(s)$ is a Markov chain on the finite state-space
$\{0,1,\ldots,N\}$. The threshold $s^*$ relates to the limiting ($t
\to \infty$ then $N \to \infty$) marginal distribution of an arbitrary
$x_k$. The probability that a randomly chosen $x_k$-value is less than
$s$ is $E[ N_t(s)] /N$ (where $E$ denotes expected value). Thus a
natural way to define a threshold $s^*$ is
\[
s^* = \sup \{ s \geq 0 : \lim_{N \to \infty}
\lim_{t \to \infty} N^{-1} E [ N_t(s)] = 0 \};
\]
the $t$-limit exists by Markov chain limit theory and it can be shown
that the $N$-limit exists too, so that $s^*$ is well defined
\cite{GKW11}.

To evaluate $s^*$, we compute the mean drift of $N_t(s)$:
\begin{eqnarray}
\label{eq:drift} \lefteqn{E[ N_{t+1} (s) - N_t (s) \mid N_t (s) = n]}
\hspace*{2cm} && \nonumber \\
 &=& 2s - (1 + F_N (n) ) \mathbf{1} \{ n > 0 \},
\end{eqnarray}
where $F_N (n) = \sum_{k=2}^n f_N(k)$, $\mathbf{1}$ denotes an indicator
function, and an empty sum is $0$. Heuristically, for large $N$ and large $n$,
$F_N(n) \approx \alpha$ by (\ref{eq:alpha}) so that this drift is approximately
$2s -1-\alpha$, and setting this equal to zero gives (\ref{eq:thres}). One
expects that the drift being zero indicates the threshold behaviour, because a
positive (negative) drift would mean $N_t(s)$ increases (decreases). In this
argument there are several limits involved ($n, N, t$ all going to $\infty$)
that need to be handled with care. We exploit techniques from Markov process
theory, such as Foster--Lyapunov conditions \cite{FMM}, to do this: we refer to
\cite{GKW11} for the details.

In conclusion, we have indicated how the distribution $f^{\rm aBS}_N(k)$ and the
quantity $\alpha$ of (\ref{eq:alpha}) capture the build-up of correlations in
Bak--Sneppen type algorithms, the threshold behaviour of which can be analysed
exactly by considering the $N_t(s)$ Markov process on a countable state space.

The class of RDPs that we have introduced is of interest in its own right.
Numerical evidence suggests that by choosing as parameter for the RDP an
appropriate statistic ($f^{\rm aBS}_N(\, \cdot \,)$) of aBS, one can
replicate the asymptotic behaviour of aBS by the RDP, for which one can
prove rigorous results more easily. The remaining analytical challenge is to
clarify the relationship between aBS and the RDP. This involves at least two
main parts: (i) proving the existence of the distributions $f^{\rm aBS}_N$
given by (\ref{aBSf}) and of the limit $\alpha$ defined by (\ref{eq:alpha});
and (ii) determining the property of aBS that allows us to use $f^{\rm
aBS}_N(\, \cdot \,)$ instead of the conditional version $f^{\rm aBS}_N(\,
\cdot \,; x)$. If one can make precise the connection between aBS and the
RDP, one should be able to transfer rigorous results for RDPs \cite{GKW11}
to aBS. In respect to challenge (i) above, it is interesting to note that if
an explicit description of $f^{\rm aBS}_N$ could be obtained, one might be
able to obtain an explicit formula for the threshold $s^*$ via
(\ref{eq:alpha}) and (\ref{eq:thres}). Finally, we note that in the case of
the classical BS process the same arguments apply, though now $f_N$ is a
function of two variables, $f_N(k,\ell)$, $k \in \{ 2, \, \ldots\, N-1\}$,
$\ell \in \{ k+1, \, \ldots, N\}$.

\begin{acknowledgments}
MG would like to acknowledge fruitful discussions with
Gregory Berkolaiko, Jack Carr, Oliver Penrose, and Michael Wilkinson.
\end{acknowledgments}

\bibliography{gkw}

\begin{thebibliography}{10}
\bibitem [{BS1()}]{BS1}%
  \BibitemOpen
  \href@noop {} {}\bibinfo {note} {This paper has 868 citations according to
  Google Scholar accessed on 8/11/10}\BibitemShut {NoStop}%
\end{thebibliography}

\end{document}